\documentclass[aps,prl,amsmath,amssymb,twocolumn]{revtex4}
\usepackage{bm}
\usepackage{dcolumn}
\usepackage{graphicx}

\newcommand{\eq}[1]{Eq.~(\ref{#1})}
\newcommand{\beq}{\begin{equation}}
\newcommand{\eeq}{\end{equation}}
\newcommand{\bdm}{\begin{displaymath}}
\newcommand{\edm}{\end{displaymath}}

\newcommand{\be}{\begin{equation}}
\newcommand{\ee}{\end{equation}}
\newcommand{\bea}{\begin{eqnarray}}
\newcommand{\eea}{\end{eqnarray}}

\newcommand{\mb}{\begin{pmatrix}}
\newcommand{\me}{\end{pmatrix}}
\newcommand{\schr}{Schr\"odinger}

\def\mat#1{\underline{\underline{\text{#1}}}} % double underscore symbolizing a matrix

\begin{document}
\title{The Nature of Quantum States Created by One Photon Absorption: Pulsed Coherent vs. Pulsed Incoherent Light}
\author{Alex C. Han$^1$, Moshe Shapiro$^{1,2}$ and Paul Brumer$^3$}
\affiliation{
$^1$Department of Physics and Astronomy, and
$^2$Department of Chemistry, The University of British Columbia, Vancouver,
and $^3$Chemical Physics Theory Group, Department of Chemistry,
The University of Toronto, Toronto, Canada}

\begin{abstract}
We analyze electronically excited nuclear wave functions and their coherence
when subjecting a molecule to the action of natural, pulsed incoherent  solar-like light,
and to that of ultrashort coherent light assumed to have the same
center frequencies and spectral bandwidths. Specifically, we
compute the spatio-temporal dependence of the excited wave packets
and their electronic coherence for these two types
of light sources, on different electronic potential energy
surfaces.  The resultant excited state wave functions are shown to be 
qualitatively different, 
reflecting the light source from which they originated.  In addition, electronic coherence is found to decay significantly
faster for incoherent light than for coherent ultrafast excitation,
for both continuum and bound wave packets.
These results confirm that the dynamics observed in 
studies using ultrashort coherent pulses
are not relevant to naturally occurring solar-induced processes
such as photosynthesis and vision.
\end{abstract}

\maketitle

%%%%%%%%%%%%%%%%
%%%%%%%%%%%%%%%%
\section*{Introduction}
%\noindent {\bf Introduction.}
An increasing number of studies in recent years have identified
surprisingly long-lived  coherences in biological systems,
especially in photosynthesic light harvesting molecules
 \cite{photosyn1,photosyn2,photosyn3,photosyn4,photosyn5,Pachon1}.
Such long-lived  coherences were detected in studies
such as two-colour photon echo spectroscopy \cite{photosyn2, photosyn4},
angle-resolved coherent optical wave-mixing \cite{photosyn3},
and phase-stabilized 2D electronic spectroscopy \cite{2Despec}.
Following these findings, it has been suggested that quantum coherence may play
an important role in biological processes, for example in the proton and
energy transfer that follow the photo-excitation step.
A challenging question is to what extent the dynamics observed in such
experiments, which utilize
coherent laser light sources, are relevant for processes induced by
incoherent  sources such as sunlight,
even when the two types of light
share the same center frequencies and spectral bandwidths.

Several previous studies have analyzed various effects of
incoherent excitation sources on the resultant molecular dynamics.
These  include an  analysis of the excited state 
survival probability as a function of the incoherence of the light \cite{jiang-brumer},
effects of incoherent fields on the photoisomerization
yields \cite{hoki-brumer}, a quantum-optical formulation of the
state of the molecule after photon-absorption \cite{paul-moshe},
studies of open system dynamics relevant to photosynthetic complexes \cite{valkunas}, 
and general features of the response of open systems to incoherent excitation \cite{Pachon2}.
However, none of these studies looked directly at the spatial shape
of the wave packets and the associated temporal evolution and coherence 
properties resulting from
excitation with incoherent vs. coherent light sources.  
Such studies are particularly important in comparing the nature of the excited quantum state 
prepared by these light sources.

In this paper we concentrate on the actual spatio-temporal
shape of the wave functions resulting from photo-excitation, a primary entity
that effectively brings out the differences between the two types of light-induced excitations mentioned above. In addition,
we consider the photo-excitation of a model molecule into a
superposition of two electronic states, and contrast the calculated
coherences induced by incoherent solar-like light with
the outcome of excitation by an ultrashort coherent pulse
with the same center frequency and spectral bandwidth.

Our main theoretical tool is
a time-dependent quantum-theoretic calculation of
the different nuclear wave packets
resulting from the photo-excitation step.
Optical incoherence is simulated 
by introducing random jumps
in the phases and center frequencies of the excitation pulse,
with the degree of optical incoherence being controlled by
the frequency and amplitudes of these random jumps.  Both the spatial structure and time dependence of
the resultant wave packet dynamics is then analyzed for a variety of
potential energy surfaces (PES).

\vspace{1ex}

\section*{Theory of ``random'' and coherent wave packet dynamics}
%\noindent {\bf Theory of ``random'' and coherent wave packet dynamics.}
Consider first the case in which a nuclear wave packet is
photo-excited to a  single excited PES.
The molecule which is assumed to be initially in the ground electronic state is
excited by a light source, treated as a classical time-varying electric field,
to a dissociative or bound excited state.
The time-evolution of the total molecular
quantum state can be described by a (stochastic) \schr's equation, written
in the adiabatic (Born Oppenheimer) electronic basis as
\be
i\hbar\frac{d}{dt} \underline{\Psi}({\bf R},t) = \mat{H}(t)
\underline{\Psi}({\bf R},t),
\ee
where
$$
\underline{\Psi}({\bf R},t)
\equiv\mb \Psi_g({\bf R},t) \\ \Psi_x({\bf R},t) \me ,
$$
with $\Psi_{g}~ (\Psi_{x})$ being
the nuclear wave packet in the ground (excited) electronic manifold.
The total Hamiltonian matrix consists of a molecular part and an electric-dipole Matter-Radiation interaction part
\bea\label{hams}
&&\mat{H}(t)        = \mat{H}^M + \mat{H}^{MR}(t), \nonumber\\
&&\mat{H}^M    = \mb {H}_g && 0 \\ 0 && {H}_x \me =
\mb {T}_N + W_g && 0 \\ 0 && {T}_N + W_x \me, \nonumber\\
&&\mat{H}^{MR}(t) = \mb 0 && V^{*MR}(t) \\  V^{MR}(t) && 0 \me~.
\eea
The nuclear kinetic energy operator ${T}_N$ and
the ground and excited PES functions
$ W_{g,x}$ of the above depend on
the set of nuclear coordinates ${\bf R}$.
The Matter-Radiation interaction term is given as,
\beq
V^{MR}({\bf R},t) = -\left[\int \psi^{e*}_x({\bf R},q_e){\bf d}({\bf R},q_e)
\psi^e_g({\bf R},q_e)dq_e\right]
\cdot \hat{\epsilon}E(t),
\label{D}
\eeq
where $q_e$ is the set of electronic coordinates,
$E(t)\equiv{\cal E}_L(t)\cos(\omega t+\Phi)$
is the electric field, $\textbf{d}(\textbf{R},q_e)$ is the dipole moment,
and $\hat{\epsilon}$ is the unit polarization vector.

It is most convenient to solve the Schr\"odinger equation in the interaction
representation (with the ${\bf R}$ variable
temporarily suppressed for brevity)
\be
i\hbar\frac{d}{dt} \bigg[\mat{U}^\dagger(t-t_0)\cdot\underline{\Psi}(t_0)
\bigg] = \mat{H}_I^{MR}(t)\cdot\bigg[\mat{U}^\dagger(t-t_0)
\cdot\underline{\Psi}(t_0) \bigg],
\ee
where $t_0$ represents some initial time long before the onset of
$\mat{H}^{MR}(t),$ and
$$
\mat{H}_I^{MR}(t) \equiv \mat{U}^\dagger(t-t_0)\mat{H}^{MR}(t)
\mat{U}(t-t_0)~,~~~~~~~~~~~~~~~~~~~~~~~~~
$$
with the (interaction representation) evolution operators defined as
\beq
 \mat{U}(t_f-t_i) \equiv e^{-{i\over\hbar}\mat{H}^M(t_f-t_i)}
= \mb {U}_g(t_f-t_i) && 0 \\ 0 && {U}_x(t_f-t_i)\me.
\eeq

Assuming the laser field intensity is weak enough so
that first-order perturbation
theory is valid, the state in the interaction representation assumes the form
\be
\mat{U}^\dagger(t-t_0)  \cdot\underline{\Psi}(t) =
\underline{\Psi}(t_0)
+\frac{1}{i\hbar}\int_{t_0}^t \mat{H}_I^{MR}(t') \cdot\underline{\Psi}(t_0) dt',
\ee
or in the \schr's representation,
\bea
\label{perturbation}
&& \underline{\Psi}(t) =
\mat{U}(t-t_0) \cdot\bigg[  1 +  \frac{1}{i\hbar}
 \int_{t_0}^t  \mat{H}_I^{MR}(t)  dt'  \bigg]
\cdot\underline{\Psi}(t_0) \nonumber\\
\eea

In order to gain physical insight we now consider
excitation by a $E(\tau)\delta(t-\tau)$ pulse,
for which $V^{MR}({\bf R},t) = D({\bf R})E(\tau)\delta(t-\tau),$
where
$$D({\bf R})\equiv -\hat{\epsilon}\cdot \int \psi^{e*}_x({\bf R},q_e){\bf d}({\bf R},q_e)
\psi^e_g({\bf R},q_e)dq_e.$$
The assumption that the molecule is initially in the ground
electronic manifold,
$
\underline{\Psi}({\bf R},t_0)  =  \mb \Psi_g({\bf R}, t_0) \\ 0 \me,
$
reduces Eq. ($\ref{perturbation}$) to
\be\label{excitedWP}
\Psi_x({\bf R},t) = {h(t-\tau)\over i\hbar}
{U}_x(t-\tau) D({\bf R})E(\tau)
{U}_g(\tau-t_0) \Psi_g({\bf R},t_0),
\ee
where
%\bea
%&{1\over 2}~{\rm for}&~ t=\tau,\nonumber\\
%h(t-\tau)~=~~&1~{\rm for}&~ t> \tau,\nonumber\\
%& 0~{\rm for}&~ t< \tau~,\nonumber
%\eea
\be
h(t-\tau) = \left\{
\begin{array}{l l}
1~{\rm for}&~ t> \tau\\
{1\over 2}~{\rm for}&~ t=\tau\\
0~{\rm for}&~ t< \tau\\
\end{array}\right.
\ee
is the Heaviside function.
The physical picture that emerges is that
no population is excited before the arrival of the delta pulse, and
that exactly at $t=\tau$, where $h(0)=1/2$, and ${U}_{e}(\tau-\tau)=1$, we have that
\bea
\Psi_x({\bf R},\tau) &=& {D({\bf R}) \over 2i\hbar }
E(\tau) {U}_g(\tau-t_0) \Psi_g({\bf R},t_0)\nonumber\\
&=& {D({\bf R})\over 2i\hbar }
E(\tau)\Psi_g({\bf R},\tau).
\eea
Thus, at $t=\tau$ we form on the excited PES
a ``replica'' of the ground state wave packet
times $D({\bf R})E(\tau)/2i\hbar,$ that evolves
at $t>\tau$ as
\be
\Psi_x({\bf R},t>\tau)=  {U}_x(t-\tau)\Psi_x({\bf R},\tau).
\ee
\begin{figure*}
\includegraphics[scale=0.75]{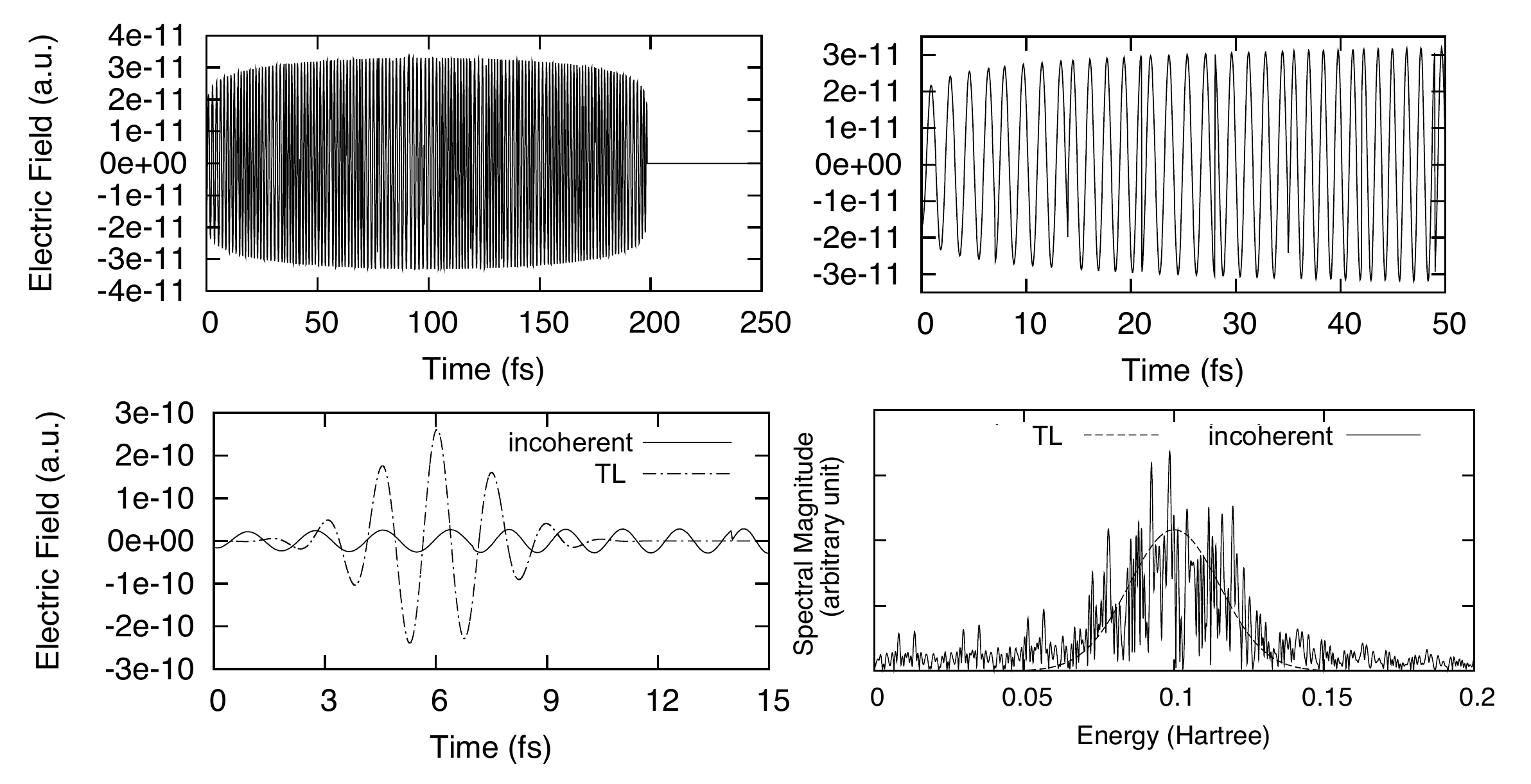}
\caption{{\bf Upper left panel} - a sample realization of incoherent light.
{\bf Upper right panel} - magnified view.
The center wavelength is $\sim$455 nm (corresponding to center energy of 0.1 Hartree). The
pulse has envelope function
${\cal E}_L(t) \sim [\sin(\pi t/200~\text{fs})]^{0.1}$ for $0<t<200$ fs,
and is zero elsewhere. The incoherent light simulates solar radiation by
introducing, every 7 fs on the average, a random phase jump in the
$[-\pi,\pi]$ range,
and a center frequency shift in the $\pm 0.0175$ Hartrees range.
{\bf Lower left panel} - A coherent
pulse whose envelope is
${\cal E}_L(t) \sim \exp\big[{-(t-6~\text{fs})^2}/{2(1.6~\text{fs})^2}\big]$.
{\bf Lower right panel} - The frequency profile of the pulses of coherent  and
incoherent light, demonstrating that both light sources share
the same center frequency and the same spectral bandwidth.}
\label{field}
\end{figure*}

Using the above insight we now
expand any time-dependent function for the electric field as a series of delta pulses,
\be
E(t)=\int_{-\infty}^\infty  E(\tau)\delta(t-\tau)d\tau.
\label{delta}
\ee
and generalize Eq. ($\ref{excitedWP}$) for a general pulse to obtain that
\beq\label{incoherent}
{\Psi}_x({\bf R},t)=
\int_{-\infty}^t {U_x(t-\tau)\over i\hbar} D({\bf R})E(\tau) {U}_g(\tau-t_0) \Psi_g({\bf R},t_0) d\tau.
\eeq
The excited wave packet is seen to be
a {\it sum} of replicas launched by all the delta pulses up till time $t,$
evolving according to the excited state Hamiltonian ${H}_x$ \cite{engelmetiu89,shapiro93}.
Written in differential form
\be\label{incoherentD}
i\hbar\frac{d}{dt}\Psi_x({\bf R},t) = {H}_x \Psi_x({\bf R},t) +
D({\bf R})E(t) {U}_g(t-t_0)\Psi_g({\bf R},t_0),
\ee
which is an inhomogeneous \schr's equation for the excited state nuclear
wave function $\Psi_x({\bf R},t)$ whose
source term is the field times the dipole operator times the
ground electronic state.

The great advantage of Eq. ($\ref{incoherentD}$) is that it allows us to
easily realize varying degrees of
incoherences in the light sources by introducing
random jumps in the center frequency and the carrier-envelope phase of
a select number of the $\delta$ function components of the electric field.
Specifically, we write the magnitude of each $\delta$ function
component of \eq{delta} as
\be\label{electric}
E(\tau) = {\cal E}_L(\tau)
\cos\big\{ [\omega_L+\Delta\omega(\tau)]\tau+\Phi(\tau)\big\}.
\ee
The real-valued functions $\Delta\omega(\tau)$ and $\Phi(\tau),$ which are
taken to be constant (zero) in time for coherent fields,
serve as random variables for incoherent light.

\vspace{2ex}

\section*{Results and Discussion}
%\noindent {\bf Results.}
Consider then the results of numerical simulations
by pulsed coherent and incoherent light.
Figure $\ref{field}$ displays the coherent pulse, and one specific realization of
the incoherent light pulse, in which we have chosen similar average values 
for the center frequency and
bandwidth, with rate of jumps of the center frequency shifts 
and phase interruptions modelling 
typical solar values.  The total energy flux,
$\int_{-\infty}^\infty |E(t)|^2 dt$, is kept the same for the two types
of pulses.   

As a model molecular system we consider the PES of the
H$_2^+$ molecular ion. We examined the
photo-excitation from the ground $1\sigma_{g}$ electronic state, to two
(dissociative or bound) excited states of the $1\sigma_u$ symmetry.
We assumed that these two states are decoupled from one another
in order to reduce the dynamics to that of a sum
of the individual ground to excited state transitions of
\eq{incoherentD}. Although this is a model system, the general results that emerge are 
expected to apply to a wide variety of systems.

\begin{figure*}
\includegraphics[scale=0.56]{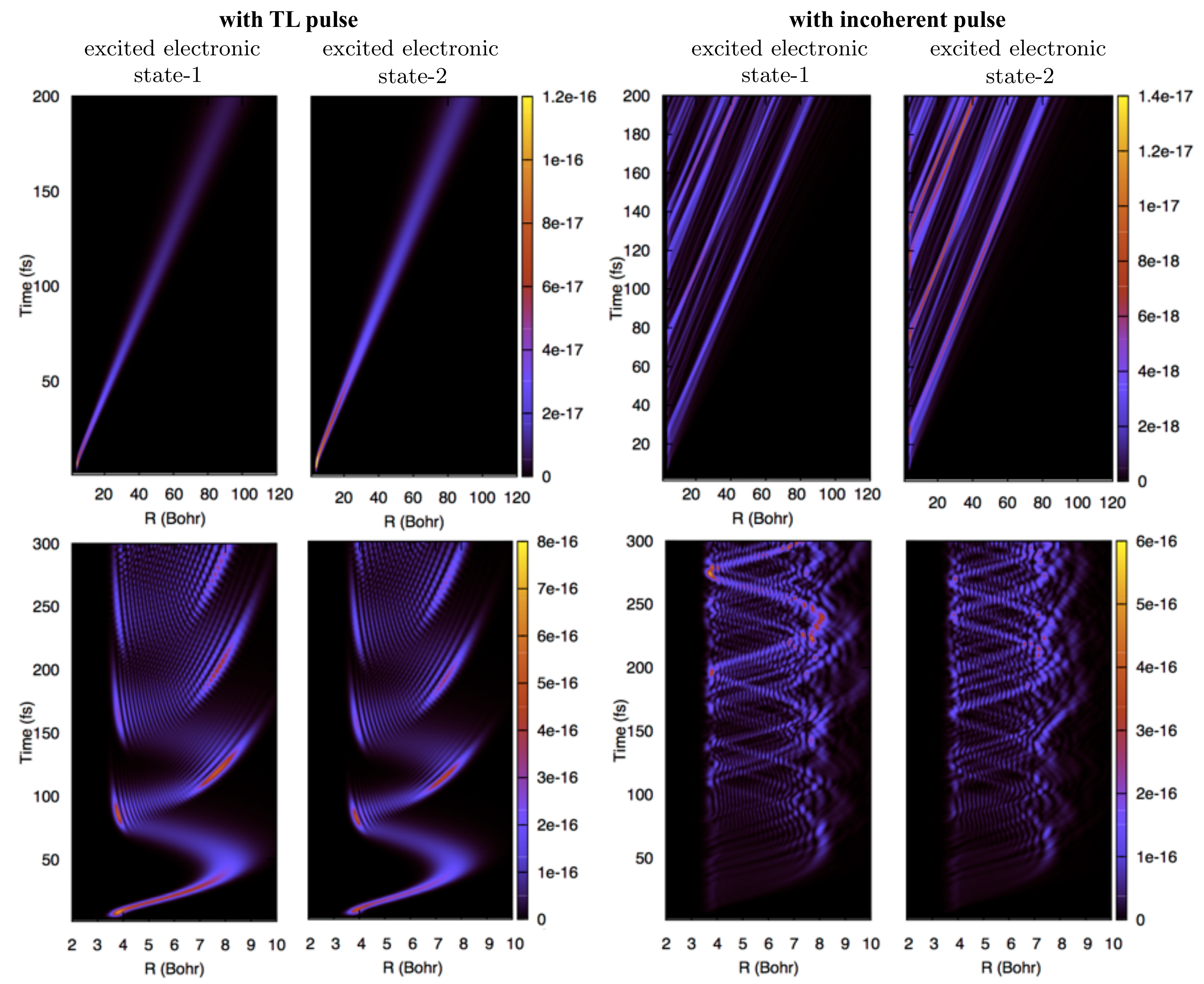}
\caption{The spatio-temporal plots of the modulus-squared 
radial wave functions of the vibrational wave packets, $u^{(1,2)}_x(R,t)$.
Top row is for purely repulsive excited state-1 and state-2 potentials, and bottom
row is for bound Morse wells. The functional forms for both types of PES correspond to the first and third situation in Figure $\ref{coherences}$.}
\label{wps-repulsive}
\end{figure*}

We assume that at $t_0 = 0$ the ground state wave function
is a ro-vibrational energy eigenstate with zero angular momentum.
The
%$\sigma_g$-to-$\sigma_u$
transition dipole function
for a linearly polarized excitation field along the $z$-axis, is given as,
$D({\bf R}) = D(R)Y_1^0(\theta)$
where
$R$ is the internuclear separation. We model $|D(R)| \approx R/2$
a.u. as in H$_2^+$ \cite{h2+edipole}.

Writing the $J=1$ wave function
as a product of vibrational and angular parts
${\Psi}_x({\bf R},t)=Y_1^0(\theta,\phi)\psi_x(R,t),$
and using atomic units,
transforms \eq{incoherent} to
\be\label{vwp}
\psi_x(R,t)= \int_{0}^t {U}^{J=1}_x(t-\tau,R) D(R)E(\tau) e^{-iE_g\tau}\psi_g(R,0) d\tau.
\ee
The excited state propagator for the $J=1$ channel is given as
\be
{U}^{J=1}_x(t-\tau,R)  = \exp\left[-i{H}^{J=1}_x(R)(t-\tau)\right],
\ee
where
\bea
{H}^J_{g,x}(R) &=& \frac{-1}{2\mu}\bigg[ \frac{d^2}{dR^2}+
\frac{2}{R}\frac{d}{dR}\bigg]
+ V^J_{g,x}(R)\nonumber\\
V^J_{g,x}(R) &=& \frac{J(J+1)}{2\mu R^2}+ W_{g,x}(R),
\eea
with $\mu$ being the reduced mass associated with the $R$ coordinate.

Using Eq. ($\ref{incoherentD}$) and Eq. ($\ref{vwp}$), we have
\be
\label{vwpd}
\frac{d\psi_x(R,t)}{dt}
= -i{H}^{J=1}_x(R) \psi_x(R,t)+
D(R)E(t) e^{-iE_gt}\psi_g(R).
\ee
Finally transforming to radial wave functions
$u_{g,x}(R,t) = R\psi_{g,x}(R,t)$, we obtain the radial analog of the
inhomogeneous \schr\ equation,
\bea\label{numerix}
\frac{d}{dt} u_x(R,t) &=& -i\bigg[\frac{-1}{2\mu}\frac{d^2}{dR^2} + V^{J=1}_{x}(R)\bigg]u_x(R,t) \nonumber\\
&+&  D(R)E(t) e^{-iE_gt}u_g(R).
\eea

Modelling the electric field $E(t)$ as in Fig. $\ref{field},$
and choosing $u_g(R)$ as the $v=5$ vibrational state,
we used Eq. (\ref{numerix}) to propagate
the pair of excited radial wave functions
$u^{(1)}_x(R,t)$ and $u^{(2)}_x(R,t)$
from $t=0$ to $300$ fs, using time-steps of $0.003$ fs and
radial-steps of $0.02$ Bohrs.
We analyze the motion of the $u^{(1)}_x(R,t)$ and $u^{(2)}_x(R,t)$ pair that
has originated from the same ground electronic state
for two cases: 1) Continuum dynamics for a pair of purely
repulsive exponentially-decaying
excited potentials. 2) Bound state dynamics for a pair of Morse potentials,
where a heavier reduced mass, $\mu=5\times918$ a.u. is used.
Sample plots of the wave packets, for these two cases, subject to
the coherent and incoherent modes of excitation,
are presented in Figs. $\ref{wps-repulsive}$.
In the case of the incoherent excitation, the results of a single 
random realization is shown.

The completed wave packets resulting from
incoherent light excitation are seen to be  choppy and highly unstructured. In the case 
of the repulsive potentials, following the
excitation step, these wave packets spread out in configuration space
much more rapidly, while covering much larger extensions,
than do the coherently excited cases.
The situation is even more dramatically seen in the bound state excitation cases
where the wave packets resulting from the incoherent light excitation
exhibit a significantly smaller number of coherent oscillations
as compared to excitation by coherent pulses, and a highly irregularly 
structured wave function is seen.

\begin{figure*}
\includegraphics[scale=0.6]{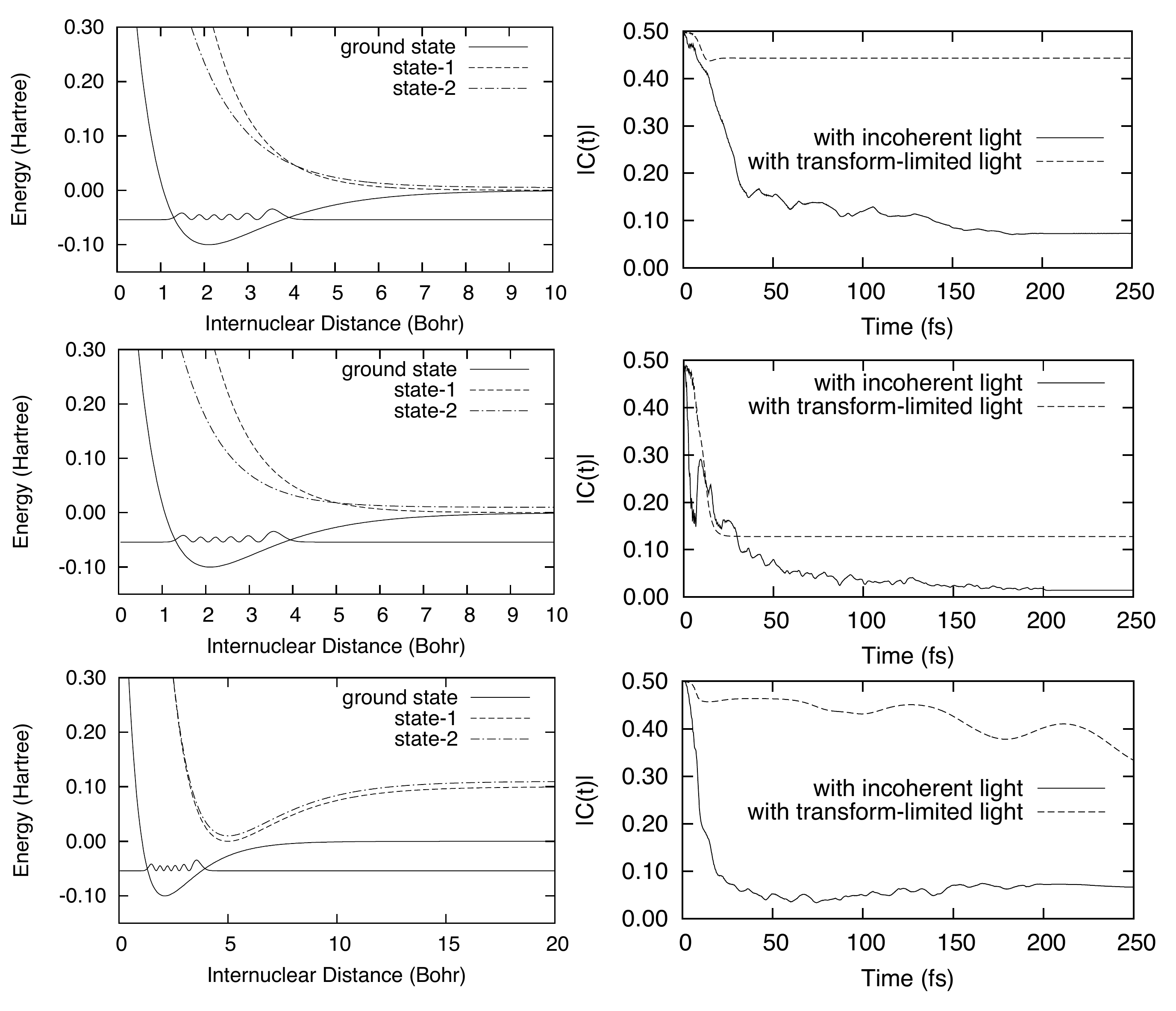}
\caption{{\bf Left column} - PES and the $v=5$ vibrational wave function.
{\bf Top panel}
Repulsive PES of the form,
$
W^{(i)}_x(R) = W^{(i)}_0 \exp[-(R-R^{(i)}_0)/a^{(i)}] +W^{(i)}_\infty,~ i=1,2;
$
where $W^{(1)}_0 = 0.1$ Hartree, $R^{(1)}_0 =  3.3$ Bohr, $a^{(1)}=1.0$ Bohr,
$W^{(1)}_\infty = 0;$ $W^{(2)}_0 = 0.1$ Hartree, $R^{(2)}_0 =  3.0$ Bohr,
$a^{(2)}=1.2$ Bohr, $W^{(2)}_\infty = 0.005$ Hartree. {\bf Middle panel}:
$W^{(1)}_0 = 0.1$ Hartree, $R^{(1)}_0 =  3.3$ Bohr, $a^{(1)}=1.0$ Bohr,
$W^{(1)}_\infty = 0$ Hartree; $W^{(2)}_0 = 0.1$ Hartree, $R^{(2)}_0 =  2.5$ Bohr,
$a^{(2)}=1.0$ Bohr, $W^{(2)}_\infty = 0.01$ Hartree.
{\bf Bottom panel}:
$
W^{(i)}_x(R) = W^{(i)}_0\big[1- e^{-(R-R^{(i)}_0)/a^{(i)}} \big]^2+W^{(i)}_\infty,~ i=1,2;
$
where $W^{(1)}_0 = 0.1$ Hartree, $R^{(1)}_0 =  5.0$ Bohr, $a^{(1)}=2.5$ Bohr,
$W^{(1)}_\infty = 0$ Hartree; $W^{(2)}_0 = 0.1$ Hartree, $R^{(2)}_0 =  5.01$ Bohr,
$a^{(2)}=2.53$ Bohr, $W^{(2)}_\infty = 0.01$ Hartree.
{\bf Right column} - The
absolute value of $\rho(t),$ the normalized electronic coherence resulting from
coherent and incoherent light
excitation for the three different PES of the left column.
The results for the incoherent light were averaged over a set of 10
random-jump realizations.}
\label{coherences}
\end{figure*}

The above results show characteristics of the excited state wave functions on individual electronic states.  
Also of interest, in particular with respect to experimental studies of long-lived coherences, is the persistence of electronic coherence, a property of several electronic levels.  To examine this we calculate the electronic coherence given by $\rho_{1,2}(t),$
the off-diagonal element of the density matrix of the two electronic states,
traced over the nuclear spatial coordinate,
\be\label{rho}
\rho_{1,2}(t)=\int u_x^{*(1)}(R,t)u_x^{(2)}(R,t) dR. 
\eeq
Because the coherence is obtained via
first-order perturbation theory, it is convenient to normalize $\rho_{1,2}(t)$ by dividing it by $\rho_{1,1}(t)+\rho_{2,2}(t)$,
the  density matrix trace, giving
\begin{equation}
C(t) = \frac{\rho_{1,2}(t)}{\rho_{1,1}(t) + \rho_{2,2}(t)}~.\label{normcoh} \end{equation}
In this way we factor out the effects of the
total excitation yield that are of no interest for this
comparative study.  Note also that Eq. (\ref{rho}), by averaging over vibrations,
includes the effect of decoherence, due to the vibrational degree of freedom, on the electronic coherences.
The resultant normalized coherences $C(t)$ are plotted in Fig. $\ref{coherences}$, for
different choices of the PES for the two electronic states,
under the action of the two types of fields.

These plots clearly show that for incoherent field excitations
the magnitude of the coherence $C(t)$ decays at a 
much faster rate than that of the  coherent pulse, approaching
a much lower asymptotic value than that of the coherent case.
This is especially so,
as shown in the middle row of Fig. $\ref{coherences}$,
when coherence $C(t)$ is not significant,  to begin with, in the coherent
case, due to small Franck-Condon  overlap regions of the two surfaces.
In the case of the bound Morse potential wells, clear and
gradual decoherence  of the two wave packets is observed for the case of 
coherent excitation. This is in strong contrast with the incoherent pulse
excitation, where the coherence $C(t)$ decays almost immediately.

In either of the three cases the final electronic coherence induced by
the incoherent field is a small fraction of the coherence generated by
the  coherent case. Thus  we find that the incoherence of the light
eliminates ``long lived coherences'' when the 
excitation is carried out with pulses of solar-like incoherent sources.

Several comments are in order. First, note that 
decoherence observed here does not include additional environmental
effects, since the molecule is isolated, i.e. the system is closed.
Rather, the observed fast coherence loss arises ,
in the incoherent light case, from the nature of the two wave packets
as they move away from one another. That is, the molecular vibration
serves as the decohering environment \cite{ignacio12}, 
and it is particularly effective
in this case due to
the highly unstructured and choppy nuclear wave functions created by incoherent
light. If, in addition, coupling to an external environment is present,
the decay of electronic coherence will also reflect this coupling \cite{Pachon2}.
As a consequence, long lived coherences are not expected to persist in realistic
open systems irradiated with incoherent sources.

Second,
note that the incoherent light source used above acts over a 200 fs time scale.  
Thus, although the 7 fs coherence time of solar radiation is represented in this 
computation, there are two significant differences between the results of this computation, 
and that which would result from solar irradiation, which acts over far longer times, and that is 
is effectively CW.  First, any pulse possesses some degree of coherence due to the pulse 
envelope \cite{jiang-brumer}.  Hence, the ``incoherent pulse" used here possesses more coherence 
than would natural solar radiation that is incident for minutes or longer.  Second, 
at such long 
times, as discussed elsewhere \cite{jiang-brumer,paul-moshe},
long time excitation of isolated molecules using natural incoherent light
leads, when averaged over realizations, to stationary eigenstates of the
Hamiltonian that do not evolve in time. 
Relaxation in open systems also
leads to mixtures of stationary states \cite{Pachon2}.
%\vspace{1ex}
\section*{Conclusions}
This paper has shown that the nature of the light-induced wave function,
as well as the rates of electronic coherence
in molecules, strongly reflect the coherence properties of the incident
radiation. Coherent light pulses produce well localized wave packets whereas 
incoherent pulses produce irregularly structured coordinate space densities.
In addition, we have calculated electronic coherences in the
photo-excitation of a molecule into a superposition of two electronic states,
using the two types of electromagnetic pulses with different degrees of
coherence. The results show decreased electronic coherence values and
much faster decoherence rates when the excitation is conducted with
pulsed incoherent light than with excitation with  coherent pulses
of the same center frequency and spectral bandwidth.
Thus, even in the absence of external environmental effects, the random character
of the incoherent light enhances decoherence.
This adds support to the conclusion \cite{paul-moshe,Pachon2,kassal13} that 
long-lived electronic coherences observed in 
biological systems are a result of the use of coherent ultrashort pulses
for molecular excitation in these experiments and that such coherences should not
be expected in natural processes induced by solar radiation.
One would anticipate that the differences noted here between coherent and
incoherent excitation would be observable in two dimensional photon echo experiments
performed with coherent vs. incoherent light \cite{turner13}.

\section*{Acknowledgement}
ACH thanks P. Xiang and E. A. Shapiro for invaluable discussions on the numerical aspects of this work. 
The work of MS was supported by the Natural Sciences and Engineering Research Council of Canada, and
that of PB by the US Air Force Office of Scientific Research under contract number FA9550-10-1-0260


\begin{thebibliography}{99}

\bibitem{photosyn1}
G. S. Engel, T. R. Calhoun, E. L. Read, T.-K. Ahn, T. Man\v{c}al, Y.-C. Cheng, R. E. Blankenship and G. R. Fleming, {\it Nature}, 2007, {\bf 446}, 782-786.
\bibitem{photosyn2}
H. Lee, Y.-C. Cheng and G. R. Fleming, {\it Science}, 2007, {\bf 316}, 1462-1465.
\bibitem{photosyn3}
I. P. Mercer, Y. C. El-Taha, N. Kajumba, J. P. Marangos, J. W. G. Tisch, M. Gabrielsen, R. J. Cogdell, E. Springate and E. Turcu, {\it Phys. Rev. Lett.},
2009, {\bf 102}, 057402.
\bibitem{photosyn4}
E. Collini, C. Y. Wong, K. E. Wilk, P. M. G. Curmi, P. Brumer and G. D. Scholes, {\it Nature}, 2010, {\bf 463}, 644-647.
\bibitem{photosyn5}
G. Panitchayangkoon, D. Hayes, K. A. Fransted, J. R. Carama, E. Harel, J. Wen, R. E. Blankenship and G. S. Engel,  {\it Proc. Natl. Acad. Sci. USA}, 2010,
{\bf 107}, 12766-12770.
\bibitem{Pachon1}
L. A. Pachon and P. Brumer, \textit{Phys. Chem. Chem. Phys}.
\bibitem{2Despec}
T. Brixner, T.  Man\v{c}al, I. V. Stiopkin and G. R. Fleming, {\it J. Chem. Phys.}, 2004, {\bf 121}, 4221-4236.
\bibitem{jiang-brumer}
X.-P. Jiang and P. Brumer, {\it J. Chem. Phys.}, 1991, {\bf 94}, 5833.
\bibitem{hoki-brumer}
H. Hoki and P. Brumer, {\it Procedia Chem.}, 2011, {\bf 3}, 122.
\bibitem{paul-moshe} P. Brumer and M. Shapiro, \textit{Proc. Natl. Acad. Sci. USA} (in press).
\bibitem{valkunas}
T. Man\v{c}al and L. Valkunas, {\it New J. Phys.}, 2010, {\bf 12}, 065044.
\bibitem{Pachon2}  L. A. Pachon and P. Brumer,\textit{ Phys. Rev. A} (submitted)  arXiv:1210.6374.  
\bibitem{h2+edipole}
A. A. Mihajlov, Lj. M. Ignjatovi\'c, N. M. Sakan, and M. S. Dimitrijevi\'c, {\it Astronomy and Astrophysics}, {\bf 469}, 2, 749, (2007)
\bibitem{engelmetiu89}
V. Engel and H.~Metiu,
{\it J. Chem. Phys.}, 1989, {\bf 90}, 6116.
\bibitem{shapiro93}
M. Shapiro, {\it J. Phys. Chem.}, 1993, {\bf 97}, 7396.
\bibitem{ignacio12} I. Franco and P. Brumer, {\it J. Chem. Phys.}, 2012 {\bf 136}, 144501. 
\bibitem{kassal13} I. Kassal, J. Yuen-Zhou and S. Rahimi-Keshari, ArXiv:1210.5022
\bibitem{turner13}
D.B. Turner, D.J. Howey, E.J. Sutor, R.A. Hensrickson, M.W. Gealy and D. J. Ulness,
ArXiv 1210.6665 


\end{thebibliography}
\end{document}